\documentclass[prb,superscriptaddress,twocolumn,showpacs,floatfix]{revtex4}
\usepackage{times}

\usepackage{graphicx} \graphicspath{{./Eps/}} \usepackage{amsmath}
\usepackage{amssymb}

\begin{document}

\title{Merging of the Dirac points in electronic artificial graphene}

\author{J. \surname{Feilhauer}} \email{juraj.feilhauer@ptb.de}

\affiliation{Physikalisch-Technische Bundesanstalt (PTB), Braunschweig,
  Germany}

\affiliation{Institute of Electrical Engineering, Slovak Academy of Sciences,
  841 04 Bratislava, Slovakia}

\author{W. \surname{Apel}}

\affiliation{Physikalisch-Technische Bundesanstalt (PTB), Braunschweig,
  Germany}

\author{L. \surname{Schweitzer}}

\affiliation{Physikalisch-Technische Bundesanstalt (PTB), Braunschweig,
  Germany}

\date{\today}

\begin{abstract}
Theory predicts that graphene under uniaxial compressive strain in an armchair direction should undergo
a topological phase transition from a semimetal into an insulator.  Due to the
change of the hopping integrals under compression, both Dirac points shift
away from the corners of the Brillouin zone towards each other. For
sufficiently large strain, the Dirac points merge and an energy gap appears.
However, such a topological phase transition has not yet been observed in
normal graphene (due to its large stiffness) neither in any other electronic
system.  We show numerically and analytically that such a merging of the Dirac
points can be observed in electronic artificial graphene created from a
two-dimensional electron gas by application of a triangular lattice of repulsive
antidots. Here, the effect of strain is modeled by tuning the distance
between the repulsive potentials along the armchair direction. Our results
show that the merging of the Dirac points should be observable in a recent
experiment with molecular graphene.
\end{abstract}


\maketitle


\section{Introduction}
In the past decade, graphene \cite{novoselov} became one of the most studied
materials due to its unusual properties that are attractive from the
perspective of basic physics as well as from that of a future technology. In
graphene, the two-dimensional hexagonal lattice of carbon  atoms provides for a
gapless electronic bandstructure which consist of two bands touching each
other at two Dirac points located at the independent corners of the hexagonal
Brillouin zone. The Fermi energy of natural graphene lies in the vicinity of
the Dirac points, where the bands have linear dispersion and conical shape.
Therefore at low energies, the bandstructure of graphene can be modeled by two
Dirac cones and electrons behave as massless Dirac particles with a valley
isospin. 

After the discovery of graphene, there has been a significant effort
\cite{artgraph, optical, photonic, photonic2, molec, molec2,
  Louie,2DEGwells,2DEGag, kalesaki} to create artificial systems with
hexagonal symmetry that mimic the unique properties of Dirac
quasiparticles. The successful experimental realization of Dirac
quasiparticles was reported in systems of cold atoms in hexagonal optical
lattices,\cite{optical} in photonic crystals created by a hexagonal array of
optical waveguides, \cite{photonic,photonic2} and in two-dimensional electron
gases with hexagonal superlattices.\cite{molec,molec2,2DEGwells,2DEGag,Louie} 
The main advantage of these
artificial systems is that one  can control and tune the system parameters
(i.e., the lattice constant and hopping parameters) independently. That makes
it possible to study features that are not observable in normal graphene,
e.g., the extremely large pseudomagnetic fields induced by strain,  Kekul\'e
lattices, edge states, etc.

One of the interesting possibilities in artificial graphene is to induce
lattice anisotropy by tuning the hopping parameters between the lattice
sites. The theory based on the nearest-neighbor tight-binding approximation
predicts \cite{pereira,hasegawa,dietl,montambaux,fuchs} that by changing the
hopping parameters, both Dirac cones are shifted away from the corners of the
Brillouin zone and become elliptical instead of circular. With increasing
anisotropy, the Dirac cones merge and an energy gap is created.  This causes
a topological phase transition \cite{fuchs,Zhu} where the system
turns from a semimetal into an insulator. At the critical point of this
merging, the energy bands touch only in a single point with an unusual
dispersion relation in its vicinity.\cite{montambaux} In one direction in
k-space, the dispersion is linear as in the standard Dirac cone but in the
perpendicular direction, the dispersion is quadratic as in the case of massive
electrons. 

In normal graphene, lattice anisotropy could be induced by application of uniaxial
strain that changes the distance between the lattice sites. But the values of
strain that would lead to the topological phase transition are unrealistically
large owing to the large graphene stiffness.\cite{pereira}  Thus,
merging of the Dirac points was demonstrated experimentally in systems of cold
atoms \cite{optical,wunsch} and microwave photonic crystals \cite{photonic}
where the distance between lattice sites can be varied much more than in
normal graphene. In spite of some efforts,\cite{molec} a merging of the Dirac
points has not yet been confirmed in any electronic system.

Here, we show analytically and numerically that a merging of Dirac points
appears also in electronic artificial graphene under uniaxial compressive
'strain'.  We study artificial graphene created in a two-dimensional electron
gas modulated by a triangular lattice of repulsive potentials, e.g., repulsive
molecules on a copper surface \cite{molec,molec2} or a muffin-tin potential of
antidots in a semiconductor heterostructure.\cite{Louie,2DEGwells,2DEGag} In
such systems, the electrons are repelled into the space between the antidots
which leads to hexagonal symmetry. The lowest energy bands of artificial graphene
are similar to the two-level tight-binding bandstructure of natural graphene
although the electrons are not bound to any attractive potential. The lattice
constant of artificial graphene can be orders of magnitude larger than in
natural graphene and, moreover, the position of the artificial lattice sites
can be easily tuned. Strain is simulated by tuning  the distance between the
repulsive triangular potentials along the armchair direction. When the antidot
lattice is stretched, the Dirac points are shifted away from the corners of the
Brillouin zone into its interior, but they always exist.   On the other hand,
with increasing compressive strain, both Dirac points are moving along the
edge of the Brillouin zone towards each other until they merge with a dispersion
as described above. For larger compression, a bulk gap is created. 

We believe that our theoretical results could be verified experimentally in
the recently created molecular graphene,\cite{molec,molec2} where a
two-dimensional electron gas on the Cu surface is modulated by a triangular
antidot lattice of repulsive molecules (CO or coronene). Here, the presence
of a Dirac energy spectrum is proven by measurements of the density of states
(DOS) which is similar to typical graphene DOS with the zero value at the
energy corresponding to the Dirac point. The authors of [\onlinecite{molec}]
performed an experiment where the artificial graphene was elongated in the
armchair direction by about 30\%. They observed no gap opening in the DOS
and that is consistent with our numerical results for the case of streched artificial
graphene. Nevertheless, our numerical calculations of the DOS show that the
molecular graphene with coronene \cite{molec2} is a very promising candidate
to observe the merging of the Dirac points and related gap opening in the DOS. 
This should occur, however, for a lattice compression of about 25\%, and that is
technologically feasible. 

This paper is organized as follows. In the second section, we review the
results of the analytical tight-binding calculations that describe the
merging of the Dirac points in the hexagonal lattice with the hopping matrix
anisotropy.  In the third section, we define mathematically the model of
artificial graphene and discuss the effect of uniaxial strain.  Then we
describe the technique that we used to calculate the numerical results. The
fourth section contains our numerical results of the bandstructure for
artificial graphene under uniaxial strain. The calculations show that the
merging of the Dirac points occurs in the artificial graphene under
compressive strain in the armchair direction. The fifth section shows the
analytical calculation of the low-energy bandstructure of strained artificial
graphene in the nearly-free electron approximation, which is valid for
weak potentials. The analytical formulas confirm the merging of the Dirac
points in the compressed artificial graphene and are in a good agreement with
the numerical results. In the sixth section, we discuss the realizability of
our calculations in real experimental situations and we conclude that the
merging of the Dirac points should be observable in the recent experiment with
molecular graphene with coronene.\cite{molec2}

\section{Merging of the Dirac points: tight-binding model description}

In this section, we review the results of the nearest-neighbor tight-binding
calculations \cite{pereira,hasegawa,dietl,montambaux,fuchs} concerning the
bandstructure of a hexagonal lattice with anisotropy in the hopping matrix
elements. We examine the positions of the Dirac points in the Brillouin zone.

\begin{figure}[t]
\centerline{\includegraphics[clip,width=0.9\linewidth]{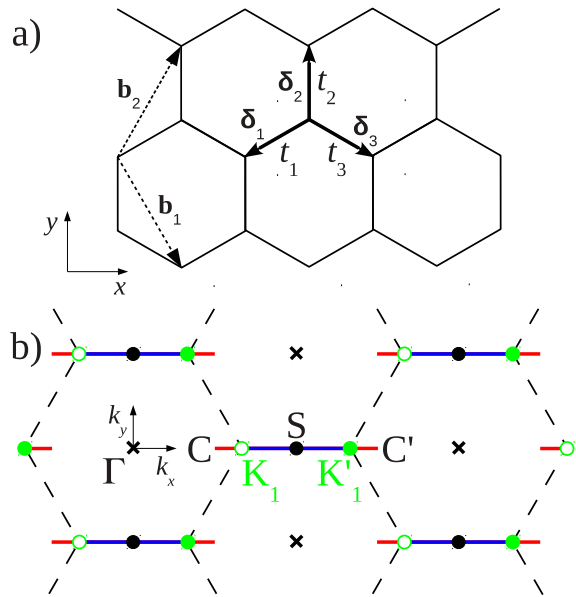}}
\vspace{-0.15cm}
\caption{(a) Model of the hexagonal lattice of graphene. Each lattice site has
  three nearest-neighbors with the locations given by the vectors $\delta_i$
  (see Eq.\ref{deltr}) and corresponding hopping parameters $t_i$. (b) The repeated
  Brillouin zone scheme of graphene. The Brillouin zone contains a
  symmetry point $\Gamma$ and a pair of independent corners K (green empty
  circles) and K' (green full circles) which are periodically repeated around
  the reciprocal lattice points (black crosses). When $t_1 = t_3 = t$ and $t_2
  = t'$, the Dirac points are located on a line $\mbox{CC}'$.  For $0 < t'/t \leq
  1$, the Dirac points are located inside the Brillouin zone (red lines) and
  for $1 < t'/t \leq 2$ the Dirac points lie at the edge (blues lines). For
  $t'/t = 2$ the Dirac points merge at S (black dots).} \label{Fig-1}
\end{figure}

Normal graphene consists of a triangular Bravais lattice with a pair of carbon
atoms located in its primitive cell. The lattice vectors are
\begin{equation}
\mathbf{b}_1 = \frac{L}{2} \mathbf{e}_x - \frac{\sqrt{3} L}{2} \mathbf{e}_y ,
\ \ \ \mathbf{b}_2 = \frac{L}{2} \mathbf{e}_x + \frac{\sqrt{3} L}{2} \mathbf{e}_y,
\label{lattv}
\end{equation}
where $L$ is a lattice constant and $L/\sqrt{3}$ is the distance between
neighboring carbon atoms. The corresponding reciprocal lattice is triangular
as well with a hexagonal Brillouin zone with the side length $4 \pi/ (3
L)$. In the nearest-neighbor tight-binding approximation, each carbon atom
possesses one localized electronic state that slightly overlaps  with its three
nearest neighbors. These overlaps are in general characterized by three
hopping parameters $t_1$, $t_2$, $t_3$ (see Fig.~\ref{Fig-1}(a)). In the
isotropic case, all hopping parameters are  equal. If we take the hopping
anisotropy in armchair direction, we get $t_1 = t_3 \neq t_2$. Then, the
bandstructure can be written in the form
\begin{equation}
\epsilon(\mathbf{k}) = \pm  |  t e^{i \mathbf{k} \cdot \boldsymbol{\delta}_1} + t'
\beta e^{i \mathbf{k} \cdot \boldsymbol{\delta}_2} + t e^{i \mathbf{k} \cdot
  \boldsymbol{\delta}_3}|,
\label{etbr}
\end{equation}
where we have defined $t_1 = t_3 = t$, $t_2 = t'$ and
\begin{equation}
\{ \boldsymbol{\delta}_1, \boldsymbol{\delta}_2, \boldsymbol{\delta}_3\} =
\frac{1}{3} \left \{-\mathbf{b}_1 - 2 \mathbf{b}_2 , \mathbf{b}_2 - \mathbf{b}_1, 2
\mathbf{b}_1 + \mathbf{b}_2 \right \},
\label{deltr}
\end{equation}
are the vectors connecting an atom with its neighbors. 

The bandstructure \eqref{etbr} consists of two bands that touch at zero energy
at the Dirac points. We obtain their position in the reciprocal space by
solving $\epsilon(\mathbf{k}) = 0$ which implies the conditions  
\begin{equation}
\cos\left( \frac{\sqrt{3}}{2} L k^D_y \right) = \pm 1, \ \ \ \cos\left(
\frac{1}{2} L k^D_x \right) = \mp \frac{t'}{2 t}.
\label{diracp}
\end{equation}
The conditions \eqref{diracp} give two independent Dirac points per Brillouin
zone which are periodically repeated through the whole reciprocal space. To
study the evolution of the Dirac points with increasing hopping anisotropy, it
is more convenient to use the repeated Brillouin zone scheme and to focus on
the Dirac points $\mbox{D}_1$ and $\mbox{D}_2$ located at
\begin{eqnarray}
k_y^{\mbox{D}_1} = k_y^{\mbox{D}_2} = 0, \\ k_x^{\mbox{D}_1} = \frac{2}{L}
\arccos\left( - \frac{t'}{2 t}\right), \\ k_x^{\mbox{D}_2} = \frac{4 \pi}{L} -
k_x^{\mbox{D}_1}.
\label{diracpp}
\end{eqnarray}
For the case of hopping anisotropy in the armchair direction, the position of
the Dirac points in reciprocal space is restricted to the $k_x$
axis. Moreover, the positions of $\mbox{D}_1$ and $\mbox{D}_2$ are symmetric
around the point S at $k_x = 2 \pi / L$.  Therefore, in the following text, we
discuss mostly only the position of $\mbox{D}_1$.

The position and existence of Dirac points is given by the value of the
anisotropy parameter $t'/t$.  Firstly, if $0 < t'/t \leq 1$, the Dirac point
$\mbox{D}_1$ lies inside the Brillouin zone (red lines in Fig.~\ref{Fig-1}(b))
and the lowest value of $k_x^{\mbox{D}_1}$ is limited by the value $\pi / L$
for $t'/t = 0$ (point C). For $t'/t = 1$ (isotropic lattice), the Dirac point
lies exactly at the corner of the Brillouin zone K with $k_x^{\mbox{D}_1} = 4
\pi / 3 L$. Secondly, if $1 < t'/t \leq 2$ the Dirac point  is located at the
edge of the Brillouin zone (blue lines in Fig.~\ref{Fig-1}(b)) and for the critical point $t'/t =
2$, both $\mbox{D}_1$ and $\mbox{D}_2$ merge at the point S. Thirdly, for $2 < t'/t$, Dirac points no longer exist and an energy
gap opens.

The hopping anisotropy discussed above could be achieved in normal graphene by
applying an uniaxial strain in the armchair direction. When the graphene
sample is stretched, the interatomic distance $|\boldsymbol{\delta}_2|$
increases. Then, the overlap of electron  wavefunctions is suppressed and the
hopping parameter $t_2$ decreases. This corresponds to the situation where
$t'/t < 1$. On the other hand, when the sample is compressed, the interatomic
distance $|\boldsymbol{\delta}_2|$ decreases and $t_2$ increases. Then $t'/t >
1$ and it increases with increasing compression.  In principle, it could be
possible to observe a merging of the Dirac points (for $t'/t = 2$) in
graphene, but, unfortunately, the mechanical stiffness of graphene is very
large \cite{pereira} and the sufficient compression to obtain $t'/t = 2$ is
unreachable.

To observe a merging of the Dirac points, one has to arrange a system with
hexagonal symmetry where it is possible to tune the hopping parameters in a
wider interval than in normal graphene. This was already shown in a lattice
of cold atoms \cite{optical,wunsch} or a photonic crystal,\cite{photonic} but
not yet for any electronic system. Our aim is to study a merging of the Dirac
points in electronic artificial graphene described in the next section.

\section{Artificial graphene under uniaxial strain}

\begin{figure}[t]
\centerline{\includegraphics[clip,width=1.0\linewidth]{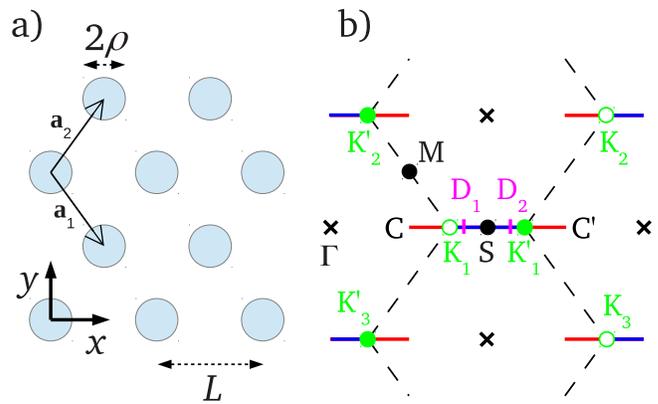}}
\vspace{-0.15cm}
\caption{(a) Triangular lattice of antidot potential spanned between the
  vectors $\mathbf{a}_1$ and $\mathbf{a}_2$. The antidot radius is $\rho/L =
  0.2$ 
  and the lattice is compressed in the $y$ (armchair) direction with $\alpha =
  1.25$. (b) The corresponding reciprocal lattice (crosses) is triangular as
  well and it is strained in the horizontal direction. In general, the first
  Brillouin zone is an irregular hexagon and we use the repeated Brillouin zone
  scheme (dashed lines). Because the antidot lattice is compressed, the Dirac
  points  $\mbox{D}_1$ and $\mbox{D}_2$ are shifted away from the corners
  $\mbox{K}_1$ and $\mbox{K}'_1$ of the Brillouin zone towards the point
  S.} \label{Fig-2}
\end{figure}

There have been several theoretical and experimental attempts to create an
electronic artificial graphene.  Most of them are based on a nanopatterning of
a two-dimensional electron gas by an external periodic potential. One
possibility is to introduce a hexagonal lattice of potential wells, which was
performed lithographically in a GaAs heterostructure.\cite{2DEGwells} In the 
present paper, we study an artificial graphene created from a two-dimensional
electron gas by application of a repulsive triangular potential of antidots
$V(\mathbf{r})$ (see Fig.~\ref{Fig-2}(a)). This kind of artificial graphene was
extensively studied \cite{Louie,tkachenko,sushkov,park,aichinger,liu} and it
captures the main features of the experiments with molecular graphene, which
is a two-dimensional electron gas on the copper surface modulated by the
repulsive molecules of CO \cite{molec} and coronene.\cite{molec2}

It can be shown \cite{Louie} that the wavefunctions of the electrons from the
two lowest energy bands are localized  around the centers of the triangles
formed by the antidots and thus form a hexagonal lattice. Consequently, the
two lowest energy bands of artificial graphene are very similar to those of
normal graphene. Namely, there is a pair of Dirac points with linear
dispersion in  their vicinity.

The effect of strain discussed in the previous section can be easily achieved
in this artificial graphene by prolonging or reducing the dimension of the
triangular antidot lattice in the armchair direction. This consequently
changes the distances in the underlying  hexagonal electron lattice.

\subsection{Trigonal antidot potential under uniaxial strain}

The antidot potential $V(\mathbf{r})$ consists of repulsive circles with radius
$\rho$ ordered on a triangular lattice with lattice constant $L$ and reads
\begin{equation}
V(\mathbf{r}) = \mathcal{V} \sum_{\mathbf{R}}  \Theta(\rho - |\mathbf{r} - \mathbf{R}|
),
\label{potko}
\end{equation}
where $\mathcal{V}$ is the potential strength, $\Theta$ is the Heaviside step
function and $\mathbf{R} = m\mathbf{a}_1 + n\mathbf{a}_2$ are the lattice sites
spanned by the vectors 
\begin{equation}
\mathbf{a}_1 = \frac{L}{2} \mathbf{e}_x - \frac{\sqrt{3} L}{2 \alpha} \mathbf{e}_y ,
\ \ \ \mathbf{a}_2 = \frac{L}{2} \mathbf{e}_x + \frac{\sqrt{3} L}{2 \alpha}
\mathbf{e}_y,
\label{vecs}
\end{equation}
where the parameter $\alpha$ represents an uniaxial 'strain' in the $y$
(armchair) direction. For $\alpha < 1$, the lattice is prolonged in the $y$
direction and for $\alpha > 1$ it is compressed. In the case $\alpha = 1$, we
obtain the unstrained triangular lattice vectors \eqref{lattv}. The
corresponding reciprocal lattice to \eqref{vecs} is $\mathbf{G} = k
\mathbf{a}^*_1 + l \mathbf{a}^*_2$, where
\begin{equation}
\mathbf{a}^*_1 = \frac{2 \pi}{L} \mathbf{e}_x - \frac{2 \pi \alpha}{\sqrt{3} L}
\mathbf{e}_y , \ \ \ \mathbf{a}^*_2 = \frac{2 \pi}{L} \mathbf{e}_x + \frac{2 \pi
  \alpha}{\sqrt{3} L} \mathbf{e}_y.
\end{equation}

The Brillouin zone of such a lattice is hexagonal (see Fig.~\ref{Fig-2}(b)), but
in contrast to the calculations in section II, it is not a regular hexagon
unless $\alpha = 1$. Its side lengths are
\begin{equation}
|\mathbf{K}_1 - \mathbf{K}'_2| = \frac{2 \pi \alpha}{\sqrt{3} L} \sqrt{1 +
  \frac{\alpha^2}{3} }, \ \ \ |\mathbf{K}_1 - \mathbf{K}'_1| = \frac{2 \pi}{L}
\left[1 - \frac{\alpha^2}{3} \right]
\label{sides}
\end{equation}
and the positions of $\mbox{K}_1$ and $\mbox{K}'_1$ points are
\begin{equation}
k_x^{\mbox{K}_1} = \frac{\pi}{L} \left[1 + \frac{\alpha^2}{3} \right],
\ \ \ k_x^{\mbox{K}'_1} = \frac{4 \pi}{L} - k_x^{\mbox{K}_1}.
\label{kbod}
\end{equation}

We would like to stress that the uniaxial constant strain applied to the
antidot lattice does not act as a constant strain on the underlying hexagonal
lattice of electrons.  For the strain $\alpha = \sqrt{3} \approx 1.73$, the
triangular antidot lattice becomes square and the corresponding 'electron'
lattice changes from hexagonal to square. A constant strain on the hexagonal
lattice would only lead to the deformation of the regular hexagons but would
not change the connectivity of the lattice. For $\alpha = \sqrt{3}$ also the
Brillouin zone is no longer hexagonal but square and we get $|\mathbf{K}_1 -
\mathbf{K}'_1| = 0$. For $\alpha > \sqrt{3}$, the antidot lattice becomes the same 
as the lattice for $\alpha \leq \sqrt{3}$ but it is rotated by 90$^{\circ}$ with
the rescaled lattice constant $L' = \sqrt{3} L / \alpha$ and strain parameter
$\alpha' = 3 / \alpha$. For example, for $\alpha = 3$, the antidot lattice again corresponds
to the unstrained artificial graphene with $L' =  L / \sqrt{3}$ and zig-zag orientation
in the $y$ direction.

The previous discussion shows, that it is sufficient to restrict the values of  $\alpha$ to the
interval $[0,\sqrt{3}]$.

\subsection{Numerical model}
The electron wavefunction in artificial graphene is described by the
Schr\"odinger equation
\begin{equation}
H \psi(\mathbf{r}) = \epsilon \psi(\mathbf{r}),
\label{sch}
\end{equation}
with the Hamiltonian
\begin{equation}
H = - \frac{\hbar^2}{2 m^*} \Delta + V(\mathbf{r}),
\label{Hsch}
\end{equation}
where the lattice potential $V(\mathbf{r})$ given by \eqref{potko} is periodic
with the lattice vectors $\mathbf{R}$. Then the electron wavefunctions are  of
the Bloch form
\begin{equation}
\psi_{\mathbf{k}}(\mathbf{r}) = \sum_{\mathbf{G}} c^{\mathbf{k}}_{\mathbf{G}} e^{i
  (\mathbf{k} + \mathbf{G} ) \cdot \mathbf{r}},
\label{psisere}
\end{equation}
where $\mathbf{k}$ is restricted to the first Brillouin zone. The Schr\"odinger
equation \eqref{sch} then becomes 
\begin{equation}
\sum_{\mathbf{G'}} \left\{ \left[ \frac{\hbar^2 (\mathbf{k} + \mathbf{G})^2}{2 m^*}
  - \epsilon \right] \delta_{\mathbf{G},\mathbf{G'}}   +  V_{\mathbf{G'}-\mathbf{G}}
\right\} c^{\mathbf{k}}_{\mathbf{G'}} = 0,
\label{schserfin}
\end{equation}
where the Fourier coefficients of the lattice potential are
\begin{equation}
V_{\mathbf{Q}} =  \frac{1}{S} \int_{\rm cell} V(\mathbf{r}) e^{i \mathbf{Q} \cdot
  \mathbf{r}} d\mathbf{r} = \frac{2 \pi \mathcal{V}}{S} \frac{\rho}{Q}
J_1(Q\rho), 
\label{fourier}
\end{equation}
where $J_1$ is the Bessel function of the first kind and $S = \sqrt{3} L^2 /
(2 \alpha)$ is the area of the unit cell.  Eq.~\eqref{fourier} shows that
due to the rotational symmetry of an antidot, the Fourier coefficients of
$V(\mathbf{r})$ depend only on the length of the wavevector $\mathbf{Q}$.
Eq.~\eqref{schserfin} shows that the periodic lattice potential mixes
only those plane waves with wavevectors that are shifted by reciprocal lattice
vectors $\mathbf{G}$.  Solving \eqref{schserfin} numerically we get the complete
band structure $\epsilon_n (\mathbf{k})$.

In general, the number of terms in the sum \eqref{psisere} is infinite (and
the number of the corresponding equations \eqref{schserfin} as well) but in
practice it is sufficient to assume a finite number of reciprocal lattice
vectors $\mathbf{G}$. Because we are interested  in the lowest energy bands
$\epsilon_n (\mathbf{k})$, the most important terms in \eqref{psisere} are the
plane waves with the smallest wavevectors. Therefore, we restrict the
summation in \eqref{psisere} to the wavevector $\mathbf{\Gamma} = \mathbf{0}$
plus a few hexagonal shells of reciprocal lattice points centered around
$\mathbf{\Gamma}$. The calculated energy bands saturate quickly with increasing
number of hexagonal shells $N$ and a sufficient value of $N$ depends on the
strength of potential. In our calculations, we use mostly $N = 20$. 

It can be easily shown that the electron eigenenergies in \eqref{schserfin}
scale as $\hbar^2/(m^*L^2)$ and therefore, we introduce the energy unit
\begin{equation}
E_0 = \frac{\hbar^2}{2 m^*} \left( \frac{4 \pi}{3 L} \right)^2,
\label{enula}
\end{equation}
which corresponds to the free electron energy at the K point. Then, we use the
dimensionless parameters $E/E_0$, $\mathcal{V}/E_0$ and $\rho/L$.

\section{Numerical results}

In this section, we present the results of our numerical calculations
describing the bandstructure of artificial graphene under uniaxial
strain. Firstly, we show the results for the unstrained case $\alpha = 1$.

\subsection{Unstrained artificial graphene}

It was shown before \cite{Louie} that the two lowest energy bands of
artificial graphene are similar to the bandstructure of normal graphene. They
touch each other at two Dirac points located at the corners K of the hexagonal
Brillouin zone. In the vicinity of the Dirac points,  the bands have linear
dispersion and form Dirac cones. 
\begin{figure}[t!]
\centerline{\includegraphics[clip,width=1.0\linewidth]{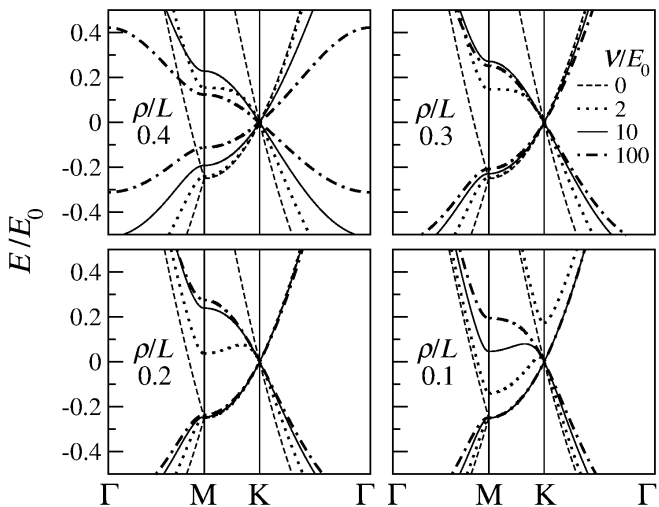}}
\vspace{-0.15cm}
\caption{Bandstructure of the unstrained ($\alpha = 1$) artificial graphene
  for various values of antidot radii $\rho$ and strengths $\mathcal{V}$ of
  the repulsive potential. The bands are shifted by a constant energy to set
  the energy at the Dirac point equal to zero.}
\label{Fig-3}
\end{figure}
Our numerical calculations are in accordance with these results as shown in
Fig.~\ref{Fig-3} for a wide range of potential parameters. The four graphs
correspond to various values of antidot radii $\rho$ and each graph contains
data for three values of potential strength $\mathcal{V}$. The free-electron
bandstructure (dashed lines) is shown for comparison. Let's focus on the two
lowest energy bands in the M-K direction. One can observe that when applying
the antidot potential,  the two lowest energy bands (which are
doubly-degenerated for $\mathcal{V} = 0$) repel each other except at the K
point where the Dirac cone is formed. This process was analytically described
in Ref.~\onlinecite{Louie} for small potential strengths in the  nearly-free
electron (NFE) approximation and its generalization for the strained
artificial graphene is presented in the following section. 

\begin{figure}[t!]
\centerline{\includegraphics[clip,width=1.0\linewidth]{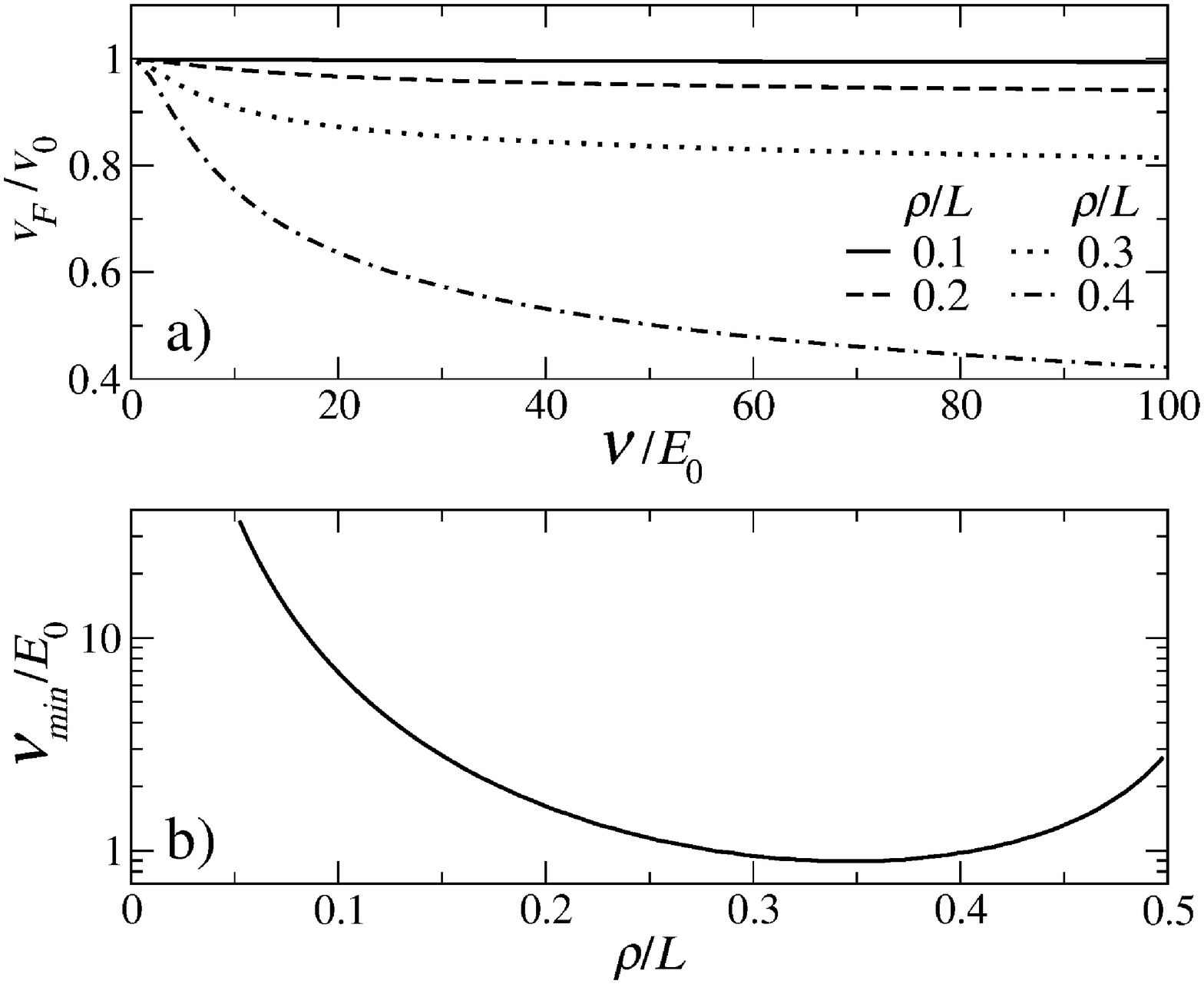}}
\vspace{-0.15cm}
\caption{(a) Ratio of the numerically calculated Fermi velocity $v_F$ and the 
  NFE value $v_0$ as a function of the potential strength for various antidot
  radii $\rho/L$. (b) Minimum value of potential strength $\mathcal{V}_{\rm min}$
  to obtain only Dirac cone  energies around the Dirac point as a function of
  antidot radius $\rho/L$.} \label{Fig-4}
\end{figure}

Figure~\ref{Fig-4}(a) shows the ratio of the numerically calculated Fermi
velocity $v_F$ (the slope of the bands crossing at K) and the NFE prediction
\cite{Louie} $v_0 = 2 \pi \hbar / (3 m^* L)$ as a function of the potential
strength for various antidot radii.

As expected, the NFE approximation is quite accurate for small values of
the potential strength independent of $\rho/L$. With increasing potential
strength, the Fermi velocity decreases, an effect that is more pronounced for larger
antidots. In the limit $\mathcal{V}/E_0 \rightarrow \infty$, the Fermi
velocity saturates at a constant value which is relatively close to $v_0$, for
$\rho/L \leq 0.3$. 

In order to have Dirac cones as the sole energy values in the vicinity of the
Dirac points, the energy gap at the M point (see Fig.~\ref{Fig-3}) should be
large enough. This energy gap increases with increasing potential strength and
for $\mathcal{V}>\mathcal{V}_{\rm min}$, the value of the second band at the point
M is higher than the Dirac point. Then, the Dirac cones are the only energy
values around Dirac points. Fig.~\ref{Fig-4}(b) shows the minimal value of the
potential strength $\mathcal{V}_{\rm min}$ as a function of the antidot
radius. For $\rho/L > 0.1$, the minimal value of the potential is of the order of
$E_0$ but for very small $\rho/L$ the potential must be much stronger.

\subsection{Artificial graphene under uniaxial strain}
As was shown in section II, the tight-binding model on a hexagonal lattice
displays Dirac points at the corners of the Brillouin zone $\text{K}_1$ and
$\text{K}'_1$. When we apply a hopping anisotropy in the armchair direction,
the Dirac points are shifted and lie somewhere at the line $\text{C}
\text{C}'$ (see Fig.~\ref{Fig-1}(b)) depending on the value of anisotropy
parameter $t'/t$. If $t'/t < 1$ the Dirac points are located inside the
Brillouin zone and for $1 \leq t'/t \leq 2$ the Dirac points lie at the edge
of the Brillouin zone. For  $t'/t = 2$ both Dirac points merge at the point S
and for $t'/t > 2$, an energy gap appears. The aim of this subsection is to
show a similar behavior in the case of electronic artificial graphene.

In normal graphene, the hopping anisotropy is induced by applying a mechanical
strain. When the lattice is stretched (or compressed), then $t'/t < 1$ (or
$t'/t > 1$). However, to describe a realistic model of normal graphene, the
hexagonal model from section II is not sufficient because the change of the
lattice geometry due to the applied strain must be taken into account.
\cite{pereira,peeters} But the results of the realistic model \cite{pereira}
capture  essentially the same features as in the case of the undeformed
hexagonal model discussed above. 

In the case of electronic artificial graphene, the effect of strain can be
obtained by tuning the lattice  constant in the armchair direction. According
to \eqref{vecs}, the values $0 < \alpha < 1$ correspond to stretching and $1 <
\alpha$ correspond to compression. Like in normal graphene, the lattice
deformations imply the deformations of the Brillouin zone.  When $0 < \alpha <
\sqrt{3}$, the Brillouin zone is hexagonal, but the length of its sides
together with the position of the K points depend on $\alpha$ as shown in
\eqref{sides} and \eqref{kbod}. 

\begin{figure}[t!]
\centerline{\includegraphics[clip,width=1.0\linewidth]{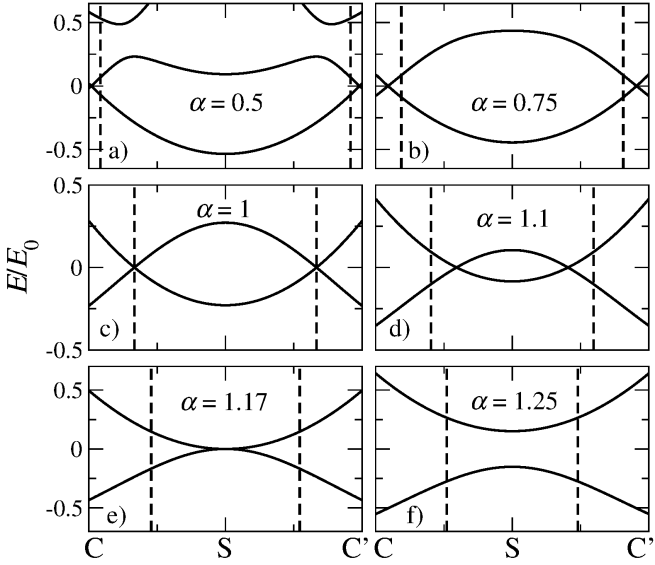}}
\vspace{-0.15cm}
\caption{Bandstructure (solid lines) of artificial graphene changing with the
  applied uniaxial strain $\alpha$. The bands correspond to the $\mbox{CC}'$
  line in the repeated Brillouin zone scheme (see Fig.~\ref{Fig-2}(b)). Figures
  (a) and (b) show the case of strained artificial graphene, figure (c)
  corresponds to the unstrained case and figures (d)-(f) depict the case of
  compressed artificial graphene. The dashed lines show the positions of
  the corners of Brillouin zone $\mbox{K}_1$ and $\mbox{K}'_1$ which change
  with applied strain according to \eqref{kbod}. The results are calculated
  for the potential strength $\mathcal{V}/E_0 = 10$ and antidot radius $\rho/L
  = 0.3$.} \label{Fig-5}
\end{figure}

Figure \ref{Fig-5} shows the bandstructure (solid lines) of artificial
graphene under uniaxial strain for potential strength $\mathcal{V}/E_0 = 10$
and antidot radius $\rho/L = 0.3$. For simplicity, we show only the bands on
the $\text{C} \text{C}'$ line which captures the positions of both Dirac
points (see Fig.~\ref{Fig-2}(b)). The graphs correspond to various values of
strain $\alpha$  and Fig.~\ref{Fig-5}(c) represents the unstrained case where
$\alpha = 1$. Here, both Dirac points lie at the K$_1$ and
K'$_1$ points depicted by dashed lines. In the case of stretching
(Fig.~\ref{Fig-5}(a), (b)), the Dirac points are located inside the Brillouin
zone 
(outside the $\text{K}_1\text{K}'_1$ edge) and with increasing stretching
(decreasing $\alpha$) the Dirac points are shifted towards the ends of
$\text{C}\text{C}'$ line. For $\alpha = 0.5$ the Dirac points almost coincide
with $\text{C}$ and $\text{C}'$. On the other hand, in the case of compressive
strain where $\alpha > 1$ (Fig.~\ref{Fig-5}(d)-(f)), the Dirac points lie at the
edge $\text{K}_1 \text{K}'_1$ and with increasing compression (increasing
$\alpha$) they approach each other. For $\alpha = 1.17$ the Dirac points merge
at the point S and the dispersion becomes parabolic as shown in 
Ref.~\onlinecite{montambaux}. For $\alpha > 1.17$, the system undergoes a
topological transition, Dirac points no longer exist and the energy gap
appears (Fig.~\ref{Fig-5}(f)).

To study the positions of the Dirac points as a function of strain, we
calculate the distance $\gamma$ of the Dirac points from the S point given by
the formula
\begin{equation}
\gamma = \frac{|\mathbf{D}_1 - \mathbf{S}|}{|\mathbf{C} - \mathbf{S}|} =
\frac{|\mathbf{D}_2 - \mathbf{S}|}{|\mathbf{C}' - \mathbf{S}|},
\end{equation}
where $|\mathbf{C} - \mathbf{S}| = |\mathbf{C}' - \mathbf{S}| = \pi/L$. If $\gamma =
1$, the Dirac points lie at $\text{C}$ and $\text{C}'$ and if $\gamma = 0$ the
Dirac points meet and merge at the S point. 

\begin{figure}[t!]
\centerline{\includegraphics[clip,width=1.0\linewidth]{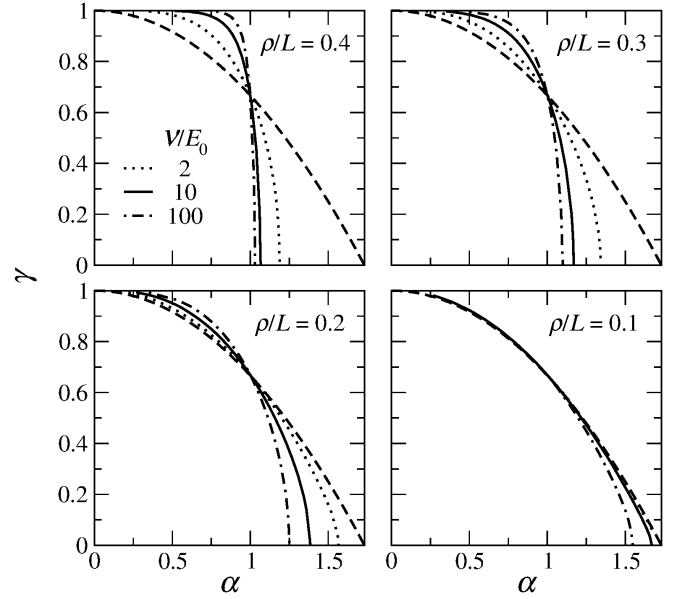}}
\vspace{-0.15cm}
\caption{Relative position of the Dirac points with respect to the S-point (see
  Fig.~\ref{Fig-2}(b)) as a function of strain. The graphs correspond to the
  various antidot radii and contain the data for different potential strengths
  and these parameters reflect those from Fig.~\ref{Fig-3}. The dashed lines
  depict the position of the K points given by \eqref{Kposi}.} \label{Fig-6}
\end{figure}

Figure \ref{Fig-6} shows the positions of the Dirac points as a function of
strain for four various antidot sizes and three potential strengths
corresponding to the parameters in Fig.~\ref{Fig-3}. The dashed lines show the
position of the points $\text{K}_1$ and $\text{K}'_1$ with respect to S given
by \eqref{kbod} as
\begin{equation}
\gamma_K = \frac{|\mathbf{K}_1 - \mathbf{S}|}{|\mathbf{C} - \mathbf{S}|} =
\frac{|\mathbf{K}'_1 - \mathbf{S}|}{|\mathbf{C}' - \mathbf{S}|} = 1 -
\frac{\alpha^2}{3}.
\label{Kposi}
\end{equation}
For $\alpha = \sqrt{3} \sim 1.73$, the points $\text{K}_1$, $\text{K}'_1$, S
merge ($\gamma_K = 0$) and the antidot lattice becomes square instead of
triangular.

Figure \ref{Fig-6} confirms that the behavior described in Fig.~\ref{Fig-5}
is universal: In the case of stretched artificial graphene ($\alpha < 1$), the
Dirac points are located inside the Brillouin zone with the  asymptotic
position at $\text{C}$ and $\text{C}'$ for $\alpha \rightarrow 0$. In the case
of zero strain ($\alpha = 1$) the Dirac points lie at $\text{K}_1$ and
$\text{K}'_1$ with $\gamma = 2/3$. When the strain is compressive ($\alpha >
1$), the Dirac points are located at the edge $\text{K}_1 \text{K}'_1$ and
move towards each other. At a critical value of strain parameter $\alpha_c$,
the Dirac points merge at S. For $\alpha > \alpha_C$ the energy gap increases
with increasing strain.

\begin{figure}[t!]
\centerline{\includegraphics[clip,width=1.0\linewidth]{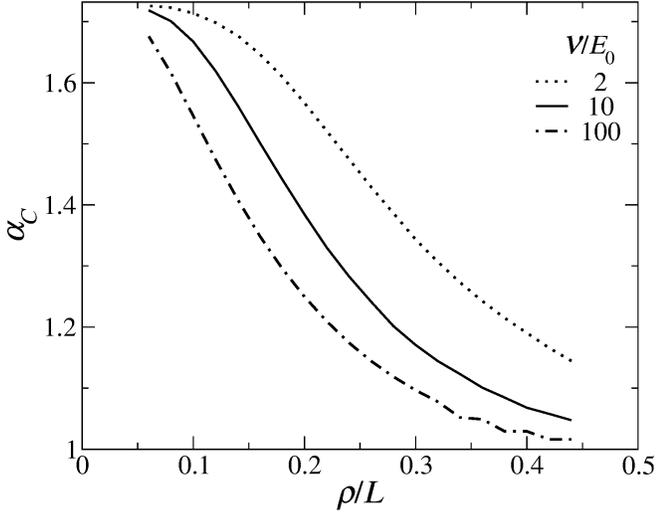}}
\vspace{-0.15cm}
\caption{Critical strain $\alpha_C$ corresponding to the merging of the Dirac
  points as a function of the antidot radius for various values of potential
  strength $\mathcal{V}$.} \label{Fig-7}
\end{figure}

Figure \ref{Fig-7} shows how $\alpha_C$ depends on the antidot radius for
various values of potential strength. For large and strong potential,
$\alpha_C$ only slightly differs from 1 which means that the artificial
graphene undergoes the topological transition  when applying very small
compressive strain. On the other hand, in the limit of small antidots, the
critical strain parameter is close to $\sqrt{3}$ for arbitrary potential
strength. This means that the Dirac points are always located at the corners
of the Brillouin zone and the system undergoes the topological transition from
a conductor into an insulator together with the transition from triangular to
square lattice. 

\section{Analytical model for weak potentials}
In this section our goal is to explain analytically the results from the
previous chapter on the basis of the NFE approach developed in Ref.
\onlinecite{Louie}. To calculate a first order correction to the
free-electron energies, the potential has to be sufficiently small.  Therefore
first we examine the validity of the NFE approach with respect to the
potential parameters.

For simplicity, we assume the unstrained artificial graphene, where the Dirac
points are located at K-points and the Brillouin zone has a shape of a regular
hexagon (the same as in Fig.~\ref{Fig-1}(b)). We are interested in the lowest
energy band around the Dirac point located at the point $\mbox{K}_1$. In the
NFE approach, the lowest-energy electron wavefunction \eqref{psisere} for
$\mathbf{k}$ close to $\mbox{K}_1$ can be approximated as a sum of three plane
waves $e^{i (\mathbf{k} - \mathbf{G}_j) \cdot \mathbf{r}}$, where  $\mathbf{G}_j =
\{\mathbf{0},\mathbf{a}^*_1,\mathbf{a}^*_2\}$ with free-electron energies close to
$E_0$ given by \eqref{enula}. The next most significant plane waves in the
expansion \eqref{psisere} would have energies close to $4E_0$. Therefore,
the NFE approach is valid when
\begin{equation}
|V_{\mathbf{G}_j}| \ll 4E_0 - E_0,
\label{podmienka}
\end{equation}
where $V_{\mathbf{G}}$ is given by \eqref{fourier}. Condition \eqref{podmienka}
can be rearranged to
\begin{equation}
\sigma \frac{4 \pi}{\sqrt{3} } \frac{1}{G_j \rho} J_1(G_j \rho) \ll 3,
\label{podm2}
\end{equation}
where
\begin{equation}
\sigma  = \frac{\mathcal{V}}{E_0} \left( \frac{\rho}{L} \right)^2
\label{sigmak}
\end{equation}
is a dimensionless antidot volume. We can replace the term $J_1(G_j \rho)/(G_j
\rho)$ in \eqref{podm2} by its maximum value $0.5$ (for $\rho/L \rightarrow
0$) and finally obtain the condition for the validity of the NFE approximation in
the form
\begin{equation}
\sigma  \ll \frac{3 \sqrt{3}}{2 \pi} \sim 1.
\label{podm3}
\end{equation}

\subsection{Dirac points}

Now we are about to derive the formula for the position $\gamma$ of the Dirac
points as a function of strain.

According to the NFE approach \cite{Louie} we can express the electron
wavefunction for $\mathbf{k}$ in the vicinity of the Dirac point $\mbox{D}_1$
(see Fig.~\ref{Fig-2}(b)) in the basis
\begin{equation}
|1\rangle = \frac{e^{i \mathbf{D_1} \cdot \mathbf{r} }}{\sqrt{S}} , \ \ \ |2\rangle =
\frac{e^{i (\mathbf{D_1} - \mathbf{a}^*_1) \cdot \mathbf{r} }}{\sqrt{S}} , \ \ \ |3\rangle
= \frac{e^{i (\mathbf{D_1} - \mathbf{a}^*_2) \cdot \mathbf{r} }}{\sqrt{S}} ,
\label{nfbasis}
\end{equation}
as
\begin{equation}
\psi(\mathbf{r}) = e^{i \mathbf{q} \cdot \mathbf{r} } \left[ b_1 |1\rangle + b_2 |2\rangle
  + b_3 |3\rangle \right],
\label{nfbas}
\end{equation}
where we have used $\mathbf{k} = \mathbf{q} + \mathbf{D}_1$ and
\begin{equation}
\mathbf{D}_1 = \frac{\mathbf{a}^*_2 + \mathbf{a}^*_1}{2} \left( 1-
\frac{\gamma}{2} 
\right)
\label{d1}
\end{equation}
is the position vector of the Dirac point $\mbox{D}_1$.  Then we can write the
Hamiltonian \eqref{Hsch} in the basis \eqref{nfbasis} as
\begin{equation}
H = H_0 + H_1,
\label{Hschnf}
\end{equation}
where
\begin{equation}
H_0 =  \left( \begin{array}{ccc} \frac{\hbar^2}{2 m^*} |\mathbf{D_1}|^2 &
  V_{\mathbf{a}^*_1} & V_{\mathbf{a}^*_2}  \\ V_{-\mathbf{a}^*_1} & \frac{\hbar^2}{2
    m^*} |\mathbf{D_1} - \mathbf{a}^*_1|^2  & V_{\mathbf{a}^*_2 - \mathbf{a}^*_1}
  \\ V_{-\mathbf{a}^*_2} & V_{\mathbf{a}^*_1 - \mathbf{a}^*_2} & \frac{\hbar^2}{2
    m^*} |\mathbf{D_1} - \mathbf{a}^*_2|^2  \end{array} \right)
\label{H0schnf}
\end{equation}
and for $qL \ll 1$
\begin{equation}
H_1 = \frac{\hbar^2}{2 m^*} \left( \begin{array}{ccc} 2 \mathbf{q} \cdot \mathbf{D_1} &
  0 & 0 \\ 0 & 2 \mathbf{q} \cdot (\mathbf{D_1} - \mathbf{a}^*_1) & 0 \\ 0 & 0 & 2
  \mathbf{q} \cdot (\mathbf{D_1} - \mathbf{a}^*_2) \end{array} \right).
\label{H1schnf}
\end{equation}
In \eqref{H0schnf}, we have omitted the diagonal terms $V_{\mathbf{0}}$ which
only cause a shift of the energy eigenvalues.  Because $|\mathbf{a}^*_1| =
|\mathbf{a}^*_2|$ and $|\mathbf{D}_1-\mathbf{a}^*_1| = |\mathbf{D}_1 - \mathbf{a}^*_2|$
the Hamiltonian $H_0$ can be written in the form
\begin{equation}
H_0 =  \left( \begin{array}{ccc} W_1 & A_1 & A_1 \\ A_1 & W_2 & A_2 \\ A_1 &
  A_2 & W_2 \end{array} \right),
\label{H0schnfred}
\end{equation}
where $A_1 = V_{\mathbf{a}^*_1} = V_{-\mathbf{a}^*_1} = V_{\mathbf{a}^*_2} =
V_{-\mathbf{a}^*_2}$, $A_2 = V_{\mathbf{a}^*_1 - \mathbf{a}^*_2} = V_{\mathbf{a}^*_2 -
  \mathbf{a}^*_1}$, $W_1 = \frac{\hbar^2}{2 m^*} |\mathbf{D_1}|^2$ and $W_2 =
\frac{\hbar^2}{2 m^*} |\mathbf{D_1} - \mathbf{a}^*_1|^2 = \frac{\hbar^2}{2 m^*}
|\mathbf{D_1} - \mathbf{a}^*_2|^2$.

We are searching for the position of the Dirac point $D_1$ and therefore we
expect that for $\mathbf{q} = 0$ two energy bands touch at $D_1$. This means
that the Hamiltonian $H(\mathbf{q} = 0) = H_0$ should have a pair of equal
eigenvalues. The eigenvalues of \eqref{H0schnfred} are
\begin{eqnarray} \label{H0eig}
\begin{split}
\epsilon_1 &= -A_2 + W_2,  \nonumber \\ \epsilon_2 &= \frac{1}{2} \bigg( A_2 +
W_1 + W_2 - \sqrt{ 8 A_1^2 + (A_2 + W_2 - W_1)^2} \bigg),  \nonumber
\\ \epsilon_3 &= \frac{1}{2} \bigg( A_2 + W_1 + W_2 + \sqrt{ 8 A_1^2 + (A_2 +
  W_2 - W_1)^2} \bigg).
\end{split}
\end{eqnarray}
We assume that the first and second eigenvalue are equal which yields the
condition
\begin{equation}
A_1^2 - A_2^2 + A_2 (W_2 - W_1) = 0.
\label{eigcon}
\end{equation}
This condition determines the position $\gamma$ of the Dirac point $D_1$ which
is hidden in the diagonal terms $W_1$ and $W_2$ in the form
\begin{equation}
W_2 - W_1 = \frac{\hbar^2}{2 m^*} \left( \frac{2 \pi}{L} \right)^2 \left[
  \gamma + \frac{\alpha^2}{3} - 1 \right].
\label{W1W2}
\end{equation}

We first explore the limit $\rho/L \ll 1$. In this case the Fourier
coefficients $V_{\mathbf{p}}$ given by \eqref{fourier} can be simplified as 
\begin{equation}
V_{\mathbf{Q}} =  \frac{2 \pi \mathcal{V}}{S} \frac{\rho}{Q} J_1(Q\rho) \sim
\frac{2 \pi \mathcal{V}}{S} \frac{\rho}{Q} \left( \frac{Q\rho}{2} \right) =
\frac{ \pi \rho^2 \mathcal{V}}{S} = V_{\mathbf{0}},
\label{fouriersim}
\end{equation}
which is independent on the wavevector $\mathbf{Q}$. Therefore $A_1 = A_2$ and
from \eqref{eigcon} we get $W_1 = W_2$. Then, from \eqref{W1W2} we get $\gamma
= 1 - \alpha^2/3$ which is exactly the same as \eqref{Kposi}. This means that
for $\rho/L \ll 1$ the Dirac point $D_1$ is located directly at the corner
$K_1$ of the  Brillouin zone. This explains the behavior observed in
Fig.~\ref{Fig-6}, where with decreasing antidot radius the position of Dirac
points is closer to the dashed line given by \eqref{Kposi}. For $\rho/L = 0.1$
and $\mathcal{V}/E_0 = 2$, the position of Dirac points and K-points is
indistinguishable. Then the critical strain $\alpha_C$, which leads to the
merging of the Dirac points, is equal to the strain corresponding to the
transition from triangular to square antidot lattice, i.e., $\alpha_C =
\sqrt{3}$.

Now we put the expressions \eqref{fourier} and \eqref{H0schnfred} into
\eqref{eigcon} and obtain $\gamma$ as a function of strain $\alpha$ and
relative barrier size $\rho/L$ in the form
\begin{eqnarray} \label{agamma}
\begin{split}
\gamma &= 1 - \frac{\alpha^2}{3} + \sigma \frac{4}{9} \frac{L}{\rho}  \Bigg\{
J_1 \left[ \frac{4 \pi \alpha}{\sqrt{3}} \frac{\rho}{L} \right] - \\ &\frac{4
  \alpha^2}{\alpha^2 + 3} \frac{J^2_1 \left[ 2 \pi \sqrt{1 +
      \frac{\alpha^2}{3}} \frac{\rho}{L} \right] }{J_1 \left[ \frac{4 \pi
      \alpha}{\sqrt{3}}  \frac{\rho}{L}  \right]} \Bigg\},
\end{split}
\end{eqnarray}
where $\sigma$ is the dimensionless volume of the barrier given by
\eqref{sigmak}.

\begin{figure}[t!]
\centerline{\includegraphics[clip,width=1.0\linewidth]{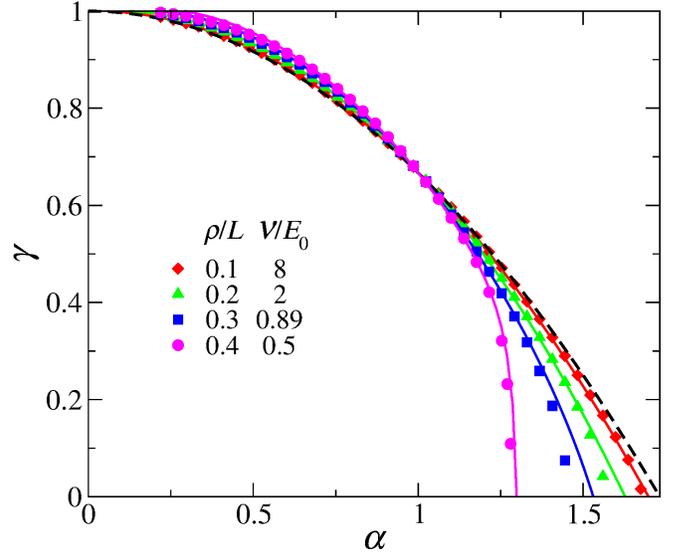}}
\vspace{-0.15cm}
\caption{Position of the Dirac points with respect to the point S as a
  function of the strain parameter for weak potentials with constant antidot
  volume $\sigma = 0.08$ and various radii listed in the legend. The 
  numerical data
  are shown as filled symbols and solid lines correspond to the analytical
  formula \eqref{agamma}. The dashed line represents the position of the
  K-point given by \eqref{Kposi}.} \label{Fig-8}
\end{figure}

Fig.~\ref{Fig-8} shows the position of the Dirac points $\gamma$ as a
function of the strain parameter $\alpha$ for weak potentials with the
antidot volume $\sigma = 0.08$. The numerical data (filled symbols) and
corresponding analytical formula \eqref{agamma}  (solid lines) are plotted for
various antidot radii shown in the legend. The analytical formula roughly
agrees with the numerical data except for the values of $\alpha$ in the
vicinity of critical value $\alpha_C$. Here, the NFE basis \eqref{nfbasis} is
not sufficient because there exists another plane wave with similar energy
(for $\mathbf{G} = \mathbf{a}^*_1 + \mathbf{a}^*_2$).  

\section{About the experimental realizability}
\begin{figure}[t!]
\centerline{\includegraphics[clip,width=1.0\linewidth]{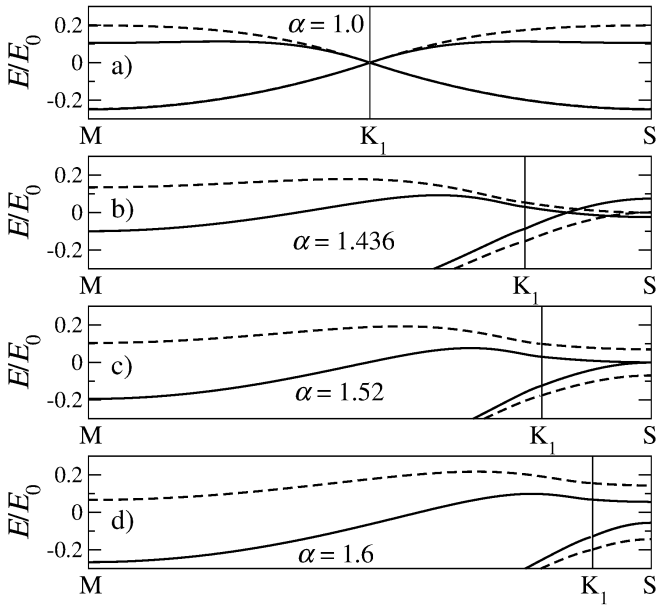}}
\vspace{-0.15cm}
\caption{The development of the bandstructure of artificial graphene with
  increasing compression given by the strain parameter $\alpha$ shown in the
  legend. The bands correspond to the line $\mbox{MK}_1\mbox{S}$ in the
  Brillouin zone (see Fig.~\ref{Fig-2}(b)) where the maximum of the first band
  and the minimum of the second band are located. The parameters are $\rho/L =
  0.2$ and $\mathcal{V}/E_0 = 3$ (solid lines) or  $\mathcal{V}/E_0 = 6$
  (dashed lines). Graph (a) shows the bands of unstrained artificial graphene
  where the Dirac cones are the only energy values around the Dirac points. Graphs
  (b) and (c) depict the merging of the Dirac point for $\mathcal{V}/E_0 = 6$
  and $\mathcal{V}/E_0 = 3$, respectively. The graph (d) shows the situation
  where the Dirac points are already merged and the gap between the bands
  exists. In the case of $\mathcal{V}/E_0 = 6$, the maximum of the first band
  is lower than the minimum of the second band, which means that the gap
  appears also in the DOS. In the case of $\mathcal{V}/E_0 = 3$, the gap
  between the bands is obscured by the second energy band and the DOS remains
  gapless.} \label{Fig-9}
\end{figure}

In experiments with artificial graphene,\cite{optical,photonic} the merging
of the Dirac points is usually revealed when measuring the density of states
(or some related quantity). When the Dirac points exist, the DOS is gapless and
equal zero at the energy of the Dirac point.  When the Dirac points merge,
the energy gap opens which is visible also in the DOS.  In this section, we
propose the experimental observation of Dirac point merging in the electronic
systems based on the artificial graphene studied in this paper.

We first discuss the conditions under which the energy gap in the DOS opens 
as a result of Dirac point merging. To observe the energy gap in the DOS, 
no electron
energies should lie in the energy window given by the gap between the
bands. Otherwise, the energy gap in the DOS would not open even if the Dirac
points merged. To analyze this problem, it is sufficient to study the
bandstructure at the edges of the Brillouin zone, because here the maximum of
the first band and minimum of the second band are located. Fig.~\ref{Fig-9}
shows how these bands change with increasing compressive strain. The
graphs represent the energy bands corresponding to the \mbox{MK$_1$S} line at
the edge of the Brillouin zone (see Fig.~\ref{Fig-2}(b)) calculated for antidot
radius $\rho/L = 0.2$ and two values of the potential strength
$\mathcal{V}/E_0 = 3$ (solid lines) and $\mathcal{V}/E_0 = 6$ (dashed lines). 

Fig.~\ref{Fig-9}(a)
shows the data for unstrained artificial graphene. According to
Fig.~\ref{Fig-4}(b), both potential strengths satisfy the condition $\mathcal{V}
> \mathcal{V}_{\rm min}$ which means that a pair of Dirac cones located at
$\mbox{K}_1$ and $\mbox{K'}_1$ (not shown here) are the sole energy values in
the vicinity  of Dirac point. With increasing compression
(Figs.~\ref{Fig-9}(b)-(d)), one can observe at the right part of each panel that
the Dirac point moves to the point S where it merges with the second Dirac
point (not shown) and the gap opens.  At the same time, the energies of the
second band in the  left part of the panels are getting lower with respect to
the Dirac point. When the Dirac points merge for $\mathcal{V}/E_0 = 3$
(Fig.~\ref{Fig-9}(c)), the energy minimum of the second band located at the M
point 
is already lower than the Dirac point. Therefore, when the gap in S opens
(Fig.~\ref{Fig-9}(d)), it does not appear in the the DOS because it is 
obscured by the second energy band. On the other hand, in the case of stronger
potential $\mathcal{V}/E_0 = 6$ when the Dirac points merge
(Fig.~\ref{Fig-9}(b)), the energies of the second band are well above the Dirac
point. When the energy gap opens (Figs.~\ref{Fig-9}(c,d)), the energies of the
second band are still higher than the maximum of the first band located at the
point S.  This means that the energy gap is visible also in the DOS.

\begin{figure}[t!]
\centerline{\includegraphics[clip,width=1.0\linewidth]{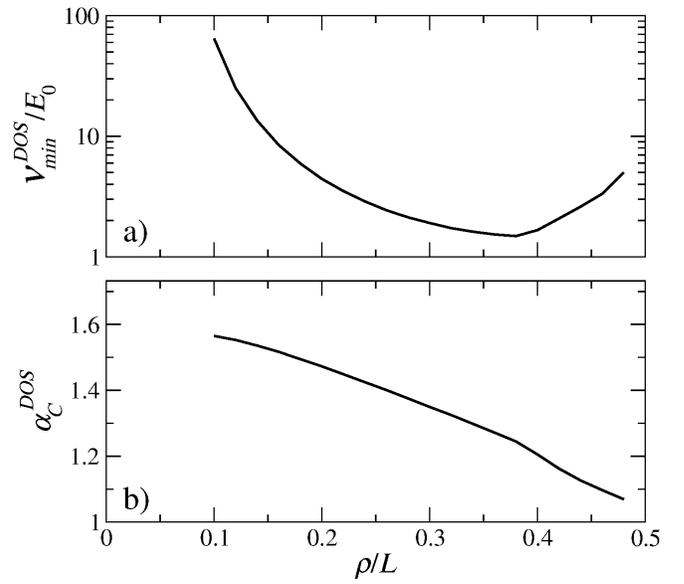}}
\vspace{-0.15cm}
\caption{(a) The minimal value of the potential strength $\mathcal{V}^{\rm
    DOS}_{\rm min}$
  that is needed to observe the gap in the DOS. This value is calculated as a
  function of the normalized antidot radius $\rho/L$. To obtain
  $\mathcal{V}^{\rm DOS}_{\rm min}$, two conditions have to be satisfied
  simultaneously. Firstly, the Dirac points merge and the bands touch
  each other only at the point S. Secondly, the minimum of the second energy
  band, which is located at the point M (see Fig.~\ref{Fig-9}), should be equal
  to the energy of the Dirac point. This means that for $\mathcal{V} =
  \mathcal{V}^{\rm DOS}_{\rm min}$ the minimum of the second energy band is equal to
  the maximum of the first band. (b) Critical strain corresponding to the
  potential with the strength $\mathcal{V}^{\rm DOS}_{\rm min}$.} \label{Fig-10}
\end{figure}

The above analysis of the bandstructure has revealed that for every $\rho/L$ there
is a minimal value of potential strength $\mathcal{V}^{\rm DOS}_{\rm min}$ that is
necessary to observe the gap opening in the DOS and this value is larger than
$\mathcal{V}_{\rm min}$ given by Fig.~\ref{Fig-4}(b).  We obtain
$\mathcal{V}^{\rm DOS}_{\rm min}$ numerically using the constraint that for critical
strain $\alpha^{\rm DOS}_C$ when the Dirac points merge, the energy value of the
second band at the point M should be equal to the energy of the Dirac point. Then,
for the potentials larger than $\mathcal{V}^{\rm DOS}_{\rm min}$ there is an interval
of strain parameter $\alpha$ where the gap in the DOS appears. For $\mathcal{V} <
\mathcal{V}^{\rm DOS}_{\rm min}$ the gapped DOS cannot be observed regardless of the
value of strain $\alpha$.  Fig.~\ref{Fig-10} shows the calculated
$\mathcal{V}^{\rm DOS}_{\rm min}$  together with the corresponding critical strain
$\alpha^{\rm DOS}_C$.

There have been a couple of attempts to observe the Dirac quasiparticles in the
experimental electronic systems based on the theoretical model of artificial
graphene studied in this paper.  In the work of Ref.~\onlinecite{2DEGag}, the
authors discussed a possible realization of artificial graphene by the
two-dimensional electron gas (2DEG) in a semiconductor heterostructure where
the triangular lattice of antidots is created by etching or gating.  Moreover,
they performed an experiment with a 2DEG made from GaAs (with the effective mass
of the electrons $m^* = 0.067 m_e$) where the antidot lattice was etched with
the lattice constant $L = 200$ nm and $\rho/L \sim 0.15$. The estimated
potential strength was about 2-4 meV.  These parameters correspond to
$\mathcal{V}/E_0 \sim 10$ where $E_0 \sim 0.25$ meV. Although the presence of
Dirac electrons was not fully confirmed in their system (due to the strong
disorder) the authors conclude that this should be technologically feasible by
reducing the lattice constant of the antidot lattice below 100 nm.

\begin{figure}[t!]
\centerline{\includegraphics[clip,width=1.0\linewidth]{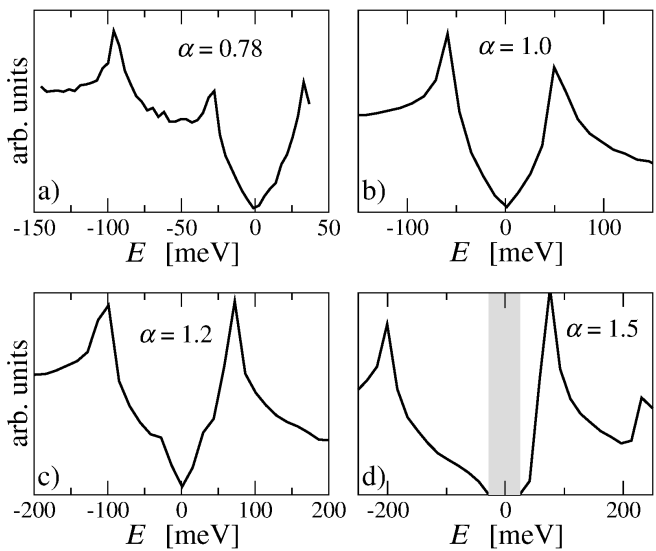}}
\vspace{-0.15cm}
\caption{The DOS of electrons in artificial graphene calculated as a function of
  the energy for various values of strain parameter $\alpha$ shown in the
  legend. The parameters $\rho = 0.7$ nm, $L = 2.7$ nm, and $\mathcal{V} = 1$ eV
  used in the simulation are the same as in the  experiment with molecular
  graphene created by coronene.\cite{molec2} (a) corresponds to the
  strained artificial graphene, (b) shows the unstrained case, and (c)-(d)
  corresponds to the artificial graphene compressed in the armchair
  direction. The gray region in (d) depicts the energy gap created as a result of
  the merging of the Dirac points.} \label{Fig-11}
\end{figure}

The same concept of artificial graphene was used in the experiments with
so-called molecular graphene \cite{molec,molec2} where the 2DEG on the copper
surface is modulated by the triangular antidot lattice of repulsive molecules
(CO or coronene). The manipulation of several hundreds of molecules is
provided by a scanning tunneling microscopy. The effective mass of surface electrons is $m^* =
0.38 m_e$ and the lattice constant of the antidot lattice $L = 1-3$ nm varied for
different experimental situations. The radius of the antidot depends on the
details of the electrostatic potential of the repulsive molecule and in the case
of coronene it is estimated \cite{molec2} to be $\rho = 0.7$ nm. In the case of
CO, the authors of Ref.~\onlinecite{molec} do not provide the exact value of
$\rho$ but it can be roughly estimated from the topographs depicting the
distribution of electron density in the modulated 2DEG as $\rho \sim 0.4$ nm.
A potential strength of an antidot was estimated roughly to be $\mathcal{V} \sim
1$ eV for both types of molecules. The presence of Dirac quasiparticles was
demonstrated by the measurements of the differential conductance between the STM
tip and the Cu surface that mimic the DOS of modulated electrons. The spatially
averaged differential conductance spectrum showed the features typical for the
DOS of graphene: a zero value at the Dirac point, linear shape around it and a
pair of van Hove singularities.

For the purposes of this paper, particularly interesting is the experiment
described in the supporting online material of the work in Ref.~\onlinecite{molec}
where the authors studied the effect of uniaxial strain on CO molecular
graphene. Here, the lattice constant was $L = 1.785$ nm which gives
$\mathcal{V}/E_0 \sim 2$ and $\rho/L \sim 0.4$. The measurement of the
differential conductance spectrum showed that when the molecular graphene is
elongated (stretched) by about 30\% in the armchair direction the DOS around
the Dirac point remains linear and gapless.  This is in agreement with our
calculations which predict that in the stretched artificial graphene the Dirac
points exist and they are shifted inside the Brillouin zone. We would like to
stress that according to our results, the merging of the Dirac points appears
if the artificial graphene is compressed (instead of stretched) in the
armchair direction. 

According to Fig.~\ref{Fig-10}(a), the potential strength $\mathcal{V}/E_0 \sim
2$ at $\rho/L \sim 0.4$ is close to the value $\mathcal{V}^{\rm DOS}_{\rm min}$. 
Therefore, it is not sure that it would be possible
to observe the merging of Dirac points in the experiment with CO
molecules. More promising is the molecular graphene made from coronene,
\cite{molec2} where the lattice constant is $L = 2.7$ nm, which gives $\rho/L
= 0.26$ and $\mathcal{V}/E_0 = 4$, a value well above
$\mathcal{V}^{\rm DOS}_{\rm min}$. Fig.~\ref{Fig-11} shows the DOS of artificial
graphene calculated for these experimental parameters. The different graphs
show how the DOS changes with strain. Figure \ref{Fig-11}(b) shows the DOS for
unstrained artificial graphene, which is very similar to the DOS of normal
graphene.  The graph \ref{Fig-11}(a) corresponds to the situation where the
artificial graphene is strained by about 30\% which models the experimental
situation discussed above. As expected, the DOS is gapless because the Dirac points
do not merge. On the other hand, the graphs \ref{Fig-11}(c-d) show how the DOS
changes with compression. The merging of the Dirac points occurs for
$\alpha_C = 1.33$, therefore for $\alpha = 1.5$ (Fig.~\ref{Fig-11}(d)), the gap
in the DOS (gray region) is nicely recognizable. The value $\alpha = 1.5$
corresponds to the compression of the antidot lattice of about
$100(1-1/\alpha) \sim 33$\% which should be possible to realize in the
experiment with molecular graphene.

As discussed in this paper, merging of the Dirac points occurs in 
artificial graphene under uniaxial compression in the armchair direction. We
would like to stress that the same effect can be obtained by stretching the
antidot lattice in the zigzag (horizontal) direction. In our model,
stretching in the zigzag direction just increases the lattice constant $L$ to
$L''$ while the vertical dimension of the antidot lattice remains the
same. Therefore, the antidot lattice stretched in zigzag direction is the same
as the lattice with lattice constant  $L''$ compressed in the armchair
direction.

\section{Conclusions}
 In normal graphene, the theory based on tight-binding approximation predicts
 that with increasing anisotropy in the hopping matrix elements, both Dirac
 points are moving along the edge of the Brillouin zone towards each other
 until they merge.  This anisotropy can be realized by application of 
 compressive strain in the armchair direction. A merging of the Dirac points
 was observed so far only in artificial systems that mimic the properties
 of graphene: in the experiments with confined microwaves in a hexagonal
 array of waveguides \cite{photonic,photonic2} and in a laser lattice with
 cold atoms.\cite{optical,wunsch} 

We have shown numerically and analytically that the merging of the Dirac
points can be observed also in  electronic artificial graphene. The
artificial graphene we considered is to be created from the two-dimensional electron gas by
applying a repulsive triangular potential and the effect of strain was modeled
by reducing the distance between the repulsive potentials along the armchair
direction.

Our numerical calculations have show that molecular graphene with coronene
\cite{molec2} is a  promising candidate to observe the merging of the Dirac
points. This should occur for a lattice compression of about 25\%, which appears
technologically feasible.

%



\begin{thebibliography}{99}
\bibitem{novoselov} A. H. Castro Neto, F. Guinea, N. M. Peres,
  K. S. Novoselov, and A. K. Geim, \emph{Rev. Mod. Phys.} {\bf 81}, 109
  (2009).

\bibitem{artgraph} M. Polini, F. Guinea, M. Lewenstein, H. C. Manoharan, and
  V. Pellegrini, \emph{Nat. Nanotechnol.} {\bf 8}, 625 (2013).

\bibitem{optical} L. Tarruell, D. Greif, T. Uehlinger, G. Jotzu, and
  T. Esslinger, \emph{Nature} {\bf 483}, 302 (2012).

\bibitem{photonic} M. Bellec, U. Kuhl, G. Montambaux, and F. Mortessagne,
  \emph{Phys. Rev. Lett.} {\bf 110}, 033902 (2013).

\bibitem{photonic2} Y. Plotnik \emph{et al.}, \emph{Nat. Mater.} {\bf 13}, 57
  (2013)

\bibitem{molec} K. K. Gomes, W. Mar, W. Ko, F. Guinea and H. C. Manoharan,
  \emph{Nature} {\bf 483}, 306 (2012) and corresponding supplemental online
  material.

\bibitem{molec2} S. Wang, L. Z. Tan, W. Wang, S. G. Louie and N. Lin,
  \emph{Phys. Rev. Lett.} {\bf 113}, 196803 (2014).

\bibitem{Louie} C.-H. Park and S. G. Louie, \emph{Nano Lett.} {\bf 9}, 1793
  (2009).

\bibitem{2DEGwells} M. Gibertini, A. Singha, V. Pellegrini, M. Polini,
  G. Vignale, A. Pinczuk, L. N. Pfeiffer, and K. W. West, \emph{Phys. Rev. B}
  {\bf 79}, 241406(R) (2009).

\bibitem{2DEGag} L. N\'advorn\'ik, M. Orlita, N. A. Goncharuk, L. Smr\v{c}ka,
  V. Nov\'ak, V. Jurka, K. Hru\v{s}ka, Z. V\'yborn\'y, Z. R. Wasilewski,
  M. Potemski, and K. V\'yborn\'y, \emph{New J. Phys.} {\bf 14}, 053002
  (2012).

\bibitem{kalesaki} E. Kalesaki, C. Delerue, C. Morais Smith, W. Beugeling,
  G. Allan, and D. Vanmaekelbergh, \emph{Phys. Rev. X} {\bf 4}, 011010 (2014).

\bibitem{hasegawa} Y. Hasegawa, R. Konno, H. Nakano, and M. Kohmoto,
  \emph{Phys. Rev. B} {\bf 74}, 033413 (2006).

\bibitem{pereira} V. M. Pereira, A. H. Castro Neto, and N. M. R. Peres,
  \emph{Phys. Rev. B} {\bf 80}, 045401 (2009).

\bibitem{dietl} P. Dietl, F. Pi\'echon, and G. Montambaux,
  \emph{Phys. Rev. Lett.} {\bf 100}, 236405 (2008).

\bibitem{fuchs} J.-N. Fuchs, arXiv:1306.0380.

\bibitem{montambaux} G. Montambaux, F. Pi\'echon, J.-N. Fuchs, and
  M. O. Goerbig, \emph{Phys. Rev. B} {\bf 80}, 153412 (2009).

\bibitem{Zhu} S.-L. Zhu, B. Wang, and L.-M. Duan, \emph{Phys. Rev. Lett.} {\bf
  98}, 260402 (2007).

\bibitem{wunsch} B. Wunsch, F. Guinea, and F. Sols, \emph{New J. Phys.} {\bf
  10}, 103027 (2008).

\bibitem{tkachenko} O. A. Tkachenko and V. A. Tkachenko, \emph{JETP Lett.}
  {\bf 99}, 204 (2014).

\bibitem{sushkov} O. P. Sushkov and A. H. Castro Neto, \emph{Phys. Rev. Lett.}
  {\bf 110}, 186601 (2013).

\bibitem{park} C.-H. Park, L. Yang, Y.-W. Son, M. L. Cohen, and S. G. Louie,
  \emph{Phys. Rev. Lett.} {\bf 101}, 126804 (2008).

\bibitem{aichinger} M. Aichinger, S. Janecek, I. Kyl\"anp\"a\"a, and
  E. R\"as\"anen, \emph{Phys. Rev. B} {\bf 89}, 235433 (2014).

\bibitem{liu} Z. Liu, J. Wang, and J. Li, \emph{Phys. Chem. Chem. Phys.} {\bf
  15}, 18855 (2013).

\bibitem{peeters} M. R. Masir, D. Moldovan, and F. M. Peeters, \emph{Solid
  State Commun.} {\bf 175-176}, 76 (2013).
\end{thebibliography}
\end{document}